\documentclass[twocolumn,showpacs,preprintnumbers,amsmath,amssymb,floatfix]{revtex4}

\usepackage{graphicx}% Include figure files
\usepackage{dcolumn}% Align table columns on decimal point
\usepackage{bm}% bold math

\newcommand{\beq}{\begin{equation}}
\newcommand{\eeq}{\end{equation}}

\begin{document}

\title{Inclusive and exclusive dilepton photoproduction at high energies}
\pacs{13.60.Fz,13.90.+i,12.40.-y,13.60.-r,13.15.Qk }
\author{C. Brenner Mariotto$^{a}$ and M.V.T. Machado$^{b}$}

\affiliation{
$^a$ Instituto de Matem\'atica, Estat\'{\i}stica e F\'{\i}sica, Universidade Federal do Rio Grande\\
Av. It\'alia, km 8, Campus Carreiros, CEP 96203-900, Rio Grande, RS, Brazil\\
$^b$ High Energy Physics Phenomenology Group, GFPAE  IF-UFRGS \\
Caixa Postal 15051, CEP 91501-970, Porto Alegre, RS, Brazil
}

\begin{abstract}
In this work we investigate the inclusive and exclusive photoproduction of dileptons,  which is relevant for the physics programme to be studied in the proposed electron-proton collider, the LHeC. In the inclusive case, the process is sensitive to the parton distribution functions in the photon whereas the exclusive  channel is connected to the small-$x$ QCD dynamics. For the latter, we investigate the role played by saturation physics at a very high energy scenario. The estimates for production cross sections and the number of events are presented. 
\end{abstract}

\pacs{13.60.Fz,13.90.+i,12.40.-y,13.60.-r,13.15.Qk}

\maketitle

\section{Introduction}

The parton distributions functions (PDFs) of the photon are fairly determined by current experiments. It is an open subject the issue of generating them from available high energy data and evolving them using perturbative QCD approach. It is timely to investigate their implications for inclusive deep inelastic processes induced by photonic parton distributions which can be tested by the forthcoming high energy beams experiments. One of them is the Deep Inelastic Electron-Nucleon Scattering at the LHC (LHeC) machine, which  is an extension of the  LHC experiment as an $ep$-collider \cite{dainton}. This proposed  extension will open a new kinematic window - the photon-proton centre-of-mass energy can reach up to TeV scale, which it is a very nice region for small-$x$ physics. Being specific, the energy of the incoming proton is delivered by the LHC beam, and a list of possible scenarios are considered for the energy of incoming electron as $E_p=7$ TeV and $E_e=50-200$ GeV, corresponding to the center of mass energies of $\sqrt{s}=2\sqrt{E_pE_e}\simeq 1.18-2.37$ TeV \cite{desreport}. The planned integrated integrated luminosity is of order $10-10^2$ fb$^{-1}$. Here, we shall consider  inclusive dilepton (Drell-Yan) production in high energy $ep$ collisions which are mostly sensitive the the quark content of photon. It is a complementary process to the heavy-quark production in $ep$ reactions that is sensitive to the gluon content of photon.

On the other hand, the exclusive photoproduction of lepton pairs, $\gamma N \rightarrow \gamma^*(\rightarrow \ell^+\ell^-)+ N$ (with $N=p,\,A$) can be studied using well established high energy factorization approaches. As an example, using simple models for the elementary dipole-hadron scattering amplitude that captures the main features of the dependence on atomic number $A$, on energy and on momentum transfer $t$ the differential cross section for exclusive dilepton photoproduction can be estimated \cite{Magno}. Such an investigation is complementary to conventional partonic description of timelike Compton scattering, which considers quark handbag diagrams at leading order in $\alpha_s$ and simple models of the relevant generalized  parton distributions (GPDs). These calculations are as well input in electromagnetic interactions in $pp$ and $AA$ collisions to be measured at the LHC. They  can also be studied with beams of
protons or anti-protons as recently done by the Tevatron machine. For instance,  the CDF Collaboration studied the exclusive production of muon pairs,
$p \overline{p} \rightarrow p \overline{p} + \mu^+ \mu^-$, at lower invariant masses \cite{CDFmu}.

The aim of this work is twofold - first, we show predictions for the photoproduction of dileptons at planned LHeC energies within high energy factorization schemes. The inclusive photoproduction cross section including the resolved and direct contributions is calculated, focusing mostly on the dilepton invariant mass distribution.  We then move to the dilepton exclusive photoproduction, analyzing the production cross section using the color dipole picture which give us a simple picture of process at small-$x$ regime. The present work is complementary to the calculation presented in Ref. \cite{MM}, where the photoproduction of massive gauge bosons was investigated as a tool to test the Standard Model and at same time to gain insight on a very specific physics beyond SM, namely the $WW\gamma $ couplings. There, the production of $W^{\pm}$, $Z$ was computed and predictions were made for the number of events produced at the LHeC regime. This article is organized as follows. The basic formulae to calculate the inclusive and exclusive (diffractive) photoproduction of virtual photons are presented in next section. Our numerical results for cross section and event rates are presented in section \ref{results}, followed by the correspondent discussion. The summary and  and conclusions are presented in section \ref{conc}.

\section{Inclusive and exclusive dilepton photoproduction}

Let start by considering the Drell-Yan inclusive cross section at next-to-leading order for the process $\gamma p\rightarrow \ell^+\ell^-+X$. In particular, we will use the NLO matrix elements and parton distribution functions (PDFs) in the $\overline{MS}$ scheme.  The differential cross section on the dilepton mass is given by \cite{BCJS},
\begin{eqnarray}
\frac{d\sigma}{dM^2} & = &  \frac{4\pi\alpha^2}{3M^2}\int_{\tau}^{1} \frac{dx_p}{x_p} \int_{\tau/x_p}^{1}\frac{dx_{\gamma}}{x_{\gamma}} \sum_{q=u,d,s,c}e_q^2 \nonumber \\
&\times & \left\{ P_{qq}(x_p,x_{\gamma},Q^2)\left[\delta \left(1-\frac{\tau}{x_px_{\gamma}} \right) \right. \right.\nonumber \\
&+& \left. \left.\frac{\alpha_s(\mu^2)}{2\pi}f_{q\bar{q}}\left(\frac{\tau}{x_px_{\gamma}} \right) \right]\right. \nonumber \\
&+& \frac{\alpha_s(\mu^2)}{2\pi}f_{gq}\left(\frac{\tau}{x_px_{\gamma}} \right)\,P_{gq}(x_p,x_{\gamma},Q^2)\,   \nonumber \\
&+& \left. \frac{6e_q^2\alpha}{2\pi}f_{gq}\left(\frac{\tau}{x_px_{\gamma}} \right)\,P_{\mathrm{cp}}(x_p,x_{\gamma},Q^2)\right\},
\label{eq:1}
\end{eqnarray}
where one has $\tau=M^2/s$. The quantities $x_{\gamma}$ and $x_p$ are the parton momentum fractions from the photon and proton, respectively. Here, $q_{\gamma},\,g_{\gamma}$ are the quark and gluon distribution functions for the photon whereas $q_{p},\,g_{p}$ are the corresponding functions for the proton.  The first term in equation above contains the dominating leading order quark-antiquark initial state and the associated real and virtual corrections. The second term includes the  $g_p+q_{\gamma}$ and $q_p+g_{\gamma}$ initial states whereas the last term is the Compton term. In the numerical calculations we set the factorization and renormalization scales at the invariant mass of the dilepton pair, $\mu^2=Q^2=M^2$. The quantities $P_i(x_p,x_{\gamma},Q^2)$ are written down as,
\begin{eqnarray}
P_{qq} & = & q_p(x_p,Q^2)\bar{q}_{\gamma}(x_{\gamma},Q^2)+ \bar{q}_p(x_p,Q^2)q_{\gamma}(x_{\gamma},Q^2),\\
P_{gq}  & = & g_p(x_p,Q^2)\left[q_{\gamma}(x_{\gamma},Q^2)+ \bar{q}_{\gamma}(x_{\gamma},Q^2)\right] \nonumber \\
&+& \left[q_{p}(x_{p},Q^2)+ \bar{q}_{p}(x_{p},Q^2)\right]g_{\gamma}(x_{\gamma},Q^2),\\
P_{\mathrm{cp}} &=& \left[q_{p}(x_{p},Q^2)+ \bar{q}_{p}(x_{p},Q^2)\right]\delta (1-x_{\gamma}).
\end{eqnarray}

The calculation above can be compared to the updated calculation for the massive gauge boson photoproduction done in Ref. \cite{MM}. They include the direct and resolved photon contributions to the production rates in an analysis designed to the LHeC machine.

As the inclusive dilepton photoproduction is concerned, a better experimental signal can be obtained considering their exclusive production. This is particularly true if one takes high photon luminosity as typical for coherent interactions in heavy ion colliders or $e^+e^-$ colliders. The high luminosity and the very large centre-of-mass energy to be available by the LHeC machine open a window for large rates of exclusive dileptons in that case. In the high energy limit it is  related to the time-like Compton scattering, where a virtual photon is produced in the photoproduction process. Such a process is better viewed in the target rest frame, commonly named the color dipole approach. Let us now to introduce the relevant equations in the color dipole picture for the timelike Compton scattering. In this formalism \cite{dipole}, the scattering process $\gamma p\rightarrow \gamma^*p$ is assumed to proceed in three stages: first the incoming real photon fluctuates into a quark--antiquark pair, then the $q\bar{q}$ pair scatters elastically on the proton, and finally the $q\bar{q}$ pair recombines to form a virtual photon (which subsequently decays into lepton pairs). The amplitude for production of the exclusive virtual photon final state is given by \cite{KMW,MPS}
\begin{eqnarray}
 \mathcal{A}^{\gamma p\rightarrow \gamma^* p}(x,Q,\Delta) & = & \sum_f \sum_{h,\bar h} \int d^2\vec{r}\,\int_0^1 d {z}\,\Psi^*_{h\bar h}(r,z,Q)\nonumber \\
& \times & \mathcal{A}_{q\bar q}(x,r,\Delta)\,\Psi_{h\bar h}(r,z,0)\,,
  \label{eq:ampvecm}
\end{eqnarray}
where $\Psi_{h\bar h}(r,z,Q)$ denotes the amplitude for a photon to fluctuate into a quark--antiquark dipole with helicities $h$ and $\bar h$ and flavor $f$. The quantity $\mathcal{A}_{q\bar q}(x,r,\Delta)$ is the elementary amplitude for the scattering of a dipole of size $r$ on the proton, $\vec{\Delta}$ denotes the transverse momentum lost by the outgoing proton (with $t=-\Delta^2$), $x$ is the Bjorken variable and $Q^2$ is the photon virtuality. Here, we are considering the photoproduction case and then only the transversely polarized overlap function contributes. For a given quark flavour $f$, when summed over the quark helicities it gives,
\begin{eqnarray}
  (\Psi_{\gamma^*}^*\Psi_{\gamma})_{T}^f & = & \frac{N_c\,\alpha_{\mathrm{em}}e_f^2}{2\pi^2}\left\{\left[z^2+\bar{z}^2\right]\varepsilon_1 K_1(\varepsilon_1 r) \varepsilon_2 K_1(\varepsilon_2 r) \right.\nonumber \\
& + &  \left. m_f^2 K_0(\varepsilon_1 r) K_0(\varepsilon_2 r)\right\},
  \label{eq:overlap_dvcs}
\end{eqnarray}
where we have defined the quantities $\varepsilon_{1,2}^2 = z\bar{z}\,Q_{1,2}^2+m_f^2$ and $\bar{z}=(1-z)$. Accordingly, the photon virtualities are $Q_1^2=0$ (incoming real photon) and $Q_2^2=-Q^2$ (outgoing virtual photon).  For simplicity, we consider space-like kinematics. In Ref. \cite{antonitcs}, the complete time-like calculation is performed in the $k_{\perp}$-formalism  and the results are somewhat larger than the presented here. Therefore, the current estimation is a lower bound for the exclusive dilepton protoproduction. 

Taking the imaginary part of amplitude, which is dominant at high energies, the elastic diffractive cross section is then given by,
\begin{eqnarray}
  \frac{d\sigma^{\gamma p\rightarrow \gamma^* p}}{d t}
  & = & \frac{1}{16\pi}\left\lvert\mathcal{A}^{\gamma p\rightarrow \gamma^* p}(x,Q,\Delta)\right\rvert^2
  \label{eq:dsdt}
\end{eqnarray}

In our numerical calculations we consider saturation models which correctly describe exclusive processes at high energies like vector meson production, diffractive DIS and DVCS. Our baseline model will be  non-forward saturation model of Ref. \cite{MPS}. Its advantage is to give the $t$ dependence of elastic differential cross section without the necessity of considerations about the impact parameter details of the process and the overall normalization is determined without  assumptions about the elastic slope. The non-forward scattering amplitude can be written as:
\begin{eqnarray}
\label{sigdipt}
\mathcal{A}_{q\bar q}(x,r,\Delta)= 2\pi R_p^2\,e^{-B|t|}N \left(rQ_{\mathrm{sat}},x\right),
\end{eqnarray}
with the asymptotic behaviors $Q_{\mathrm{sat}}^2(x,\Delta)\sim
\max(Q_0^2,\Delta^2)\,\exp[-\lambda \ln(x)]$. Specifically, the $t$ dependence of the saturation scale is parametrised as
\begin{eqnarray}
\label{qsatt}
Q_{\mathrm{sat}}^2\,(x,|t|)=Q_0^2(1+c|t|)\:\left(\frac{1}{x}\right)^{\lambda}\,, \end{eqnarray}
in order to interpolate smoothly between the small and intermediate transfer
regions. The form factor $F(\Delta)=\exp(-B|t|)$ catches the transfered momentum  dependence of the proton vertex, which is factorised from the
projectile vertices and  does not spoil the geometric scaling properties. Finally, the scaling function $N$ is obtained from the forward saturation model
\cite{Iancu:2003ge}, whose functional form resembles the analytical solution for the nonlinear QCD equations in the asymptotic energy regime. 
Here, we investigate the exclusive photoproduction of a heavy timelike photon which decays into a lepton pair, $\gamma p \rightarrow \ell^+\ell^-p$. Therefore, for the $\ell^+\ell^-$ invariant mass distribution from the virtual $\gamma^*$ decay we have (with $Q^2=M_{\ell^+\ell^-}^2$),
\begin{eqnarray}
 \frac{d\sigma }{dM_{\ell^+\ell^-}^2}\left(\gamma p\rightarrow \ell^+\ell^- p\right) = \frac{\alpha_{em}}{3\pi M_{\ell^+\ell^-}^2}\,\sigma\left(\gamma p \rightarrow \gamma^* p \right).
\label{eq:dilepton}
\end{eqnarray}

In the next section we compute the numerical results for the inclusive and exclusive photoproduction of dileptons focusing on  LHeC regime of energy/luminosity. We also investigate the  sensitivity to to the theoretical uncertainties for both processes.

\section{Numerical results and discussions}
\label{results}

\begin{figure}[t]
\includegraphics[scale=0.5]{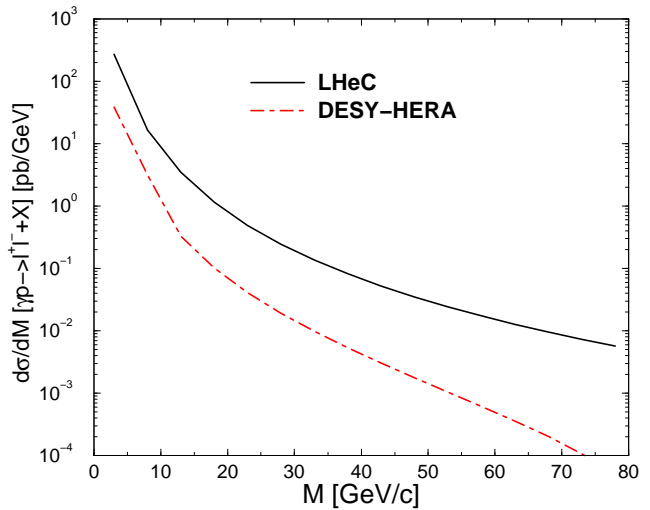}
\caption{Differential cross sections for the inclusive photoproduction of dileptons as a function of dilepton invariant mass for DESY-HERA and LHeC energies.}
\label{fig:1}
\end{figure}

Our starting point is a study of the inclusive dilepton photoproduction for the proposed LHeC machine \cite{dainton,desreport}. Using the design with a electron beam having laboratory energy of $E_e=70$ GeV, the center of mass energy will reach $E_{cm}=W_{\gamma p}=1.4$ TeV and a nominal luminosity of order $10^{33}$ cm$^{-2}$s$^{-1}$. In Fig. \ref{fig:1} is presented the differential cross section , Eq. (\ref{eq:1}), as a function of the dilepton invariant mass (with $M_{\ell^+\ell^-}>3$ GeV/c). Here, we have summed the resolved and direct contributions and through the calculations proton structure functions of CTEQ6M \cite{Pumplin:2002vw} and  photon structure functions of GRV \cite{grvphoton} have been used
with $\mu^{2}=M_{\ell^+\ell^-}^{2}$. For sake of comparison it is also shown the result for the DESY-HERA energy range. When comparing the distinct energies, deviations are larger for higher invariant masses. The behavior is similar for small invariant mass and the overall normalization differs by a factor about ten. Our estimates for the inclusive DY photoproduction cross sections are the following. One gets $\sigma_{\mathrm{inc}}(\gamma +p\rightarrow \ell^+\ell^-X)\simeq 0.78$ nb for the integrated cross section with $M_{\ell^+\ell^-}\geq 3$ GeV/c. This results can be compared to the HERA energy, which gives an integrated cross section of 0.12 nb.  These are rough estimates once we have not introduced the theoretical uncertainty coming from using distinct PDFs for proton and photons. The uncertainty concerning the variation of hard scale is small in a NLO QCD computation.

We also compute the electron-proton total cross sections for mass cut  $M_{\ell^+\ell^-}\geq 3$  GeV/c and the corresponding number of events. The latter has been computed using $N_{ev}=\sigma(e p\rightarrow \ell^+\ell^-+X)L_{int}$. At this point we consider the acceptance in the dilepton channel as 100\%.   The photoproduction cross section is calculated by convoluting the Weizs\"{a}cker-Williams spectrum
\begin{eqnarray}
f_{\gamma/e}(y)& = & \frac{\alpha}{2\pi}\left[\frac{1+(1-y)^2}{y}\log \frac{Q^2_{max}}{Q^2_{min}} \right. \nonumber \\
& - & \left.2m_e^2y\,\left(\frac{1}{Q^2_{min}}- \frac{1}{Q^2_{max}}\right)  \right],
\end{eqnarray}
with the differential hadronic cross section. One has  $Q^2_{min}=m_e^2y/(1-y)$ and we impose a cut of $Q^2_{min}=0.01$. An integrated
luminosity $L_{int}$ of 10 fb$^{-1}$ \cite{desreport} is assumed  in order to compute the number of events, $N_{ev}$. The final result is $\sigma(e p\rightarrow \ell^+\ell^-+X)=11$ pb for the LHeC energy regime and the number of events is around $1.1\times 10^5$. The number of events is considerable as expected for an inclusive process.

\begin{figure}[t]
\includegraphics[scale=0.5]{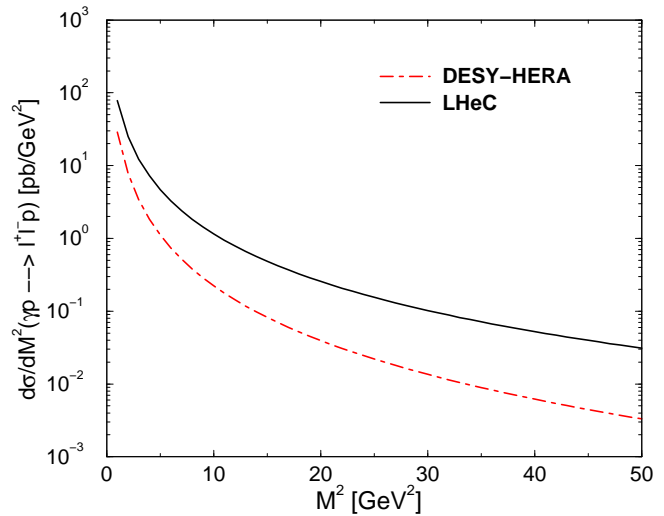}
\caption{(Color online) Differential cross sections for the exclusive photoproduction of dileptons as a function of dilepton invariant mass for DESY-HERA and LHeC energies.}
\label{fig:2}
\end{figure}

Let us now move to the exclusive dilepton photoproduction. In the following we compute the integrated cross section, performing the integration over $|t|\leq 1$ GeV$^2$. It should be noticed that the $|t|$-dependence is distinct from other implementations of the dipole cross section. However, for integrated cross sections, these deviations do not play an important role. In Fig. \ref{fig:2} we focus on the invariant mass ($M_{\ell^+\ell^-}^2\geq 1$ GeV$^2$/c$^2$)  dependence for fixed values of energy, as computed from Eqs. (\ref{eq:dsdt}) and (\ref{eq:dilepton}). Once again we show the results for the LHeC energy and DESY-HERA for comparison.  Accordingly, the spectrum is dominated by small invariant masses of order $M_{\ell^+\ell^-}^2<20$ GeV$^2$, independent of energy. In the saturation models, the qualitative behavior can be obtained considering dominance of small size dipoles configuration.  The forward amplitude reads as ${\cal A}\propto (Q_{\mathrm{sat}}^2/M^2)$ times a logarithmic enhancement on $M^2$. Therefore, qualitatively the differential cross section behaves like $d\sigma/dM^2\propto (M^2)^{-\delta}[1+\log(M^2)]$, with $\delta \approx 3$. The total cross section, $\sigma(\gamma p \rightarrow \ell^+\ell^-\,p)$, integrated over dilepton invariant mass $M_{\ell^+\ell^-}\geq 1.5 $ GeV for energies $W>>10^2$ GeV can be adjusted by a power fit $\sigma = 3\,\mathrm{pb}\,(W/W_0)^{0.46}$, with $W_0=1$ GeV. The integrated cross section gives $\sigma_{\mathrm{exc}}(\gamma +p\rightarrow \ell^+\ell^-+p)\simeq 80$ pb with the cut $M_{\ell^+\ell^-}\geq 1.5$ GeV/c and also one has  $\sigma(e p\rightarrow \ell^+\ell^-+p)=2.1$ pb for the LHeC energy regime and the number of events is around $2\times 10^4$.  The number of events is one order of magnitude smaller than for the inclusive case. This is consistent with the HERA trend of a 10\% ratio for diffractive over inclusive processes. The extrapolation of current study for the electroproduction case is straightfoward as within the color dipole approach the deeply virtual Compton cross section is in general following the geometric scaling property. That is, $\sigma_{\mathrm{Compton}} (x,Q^2) = \sigma_{\mathrm{Compton}} (\tau )$  , where the scaling variable $Q$ is given in terms of incoming photon virtuality and the corresponding saturation scale, $\tau = Q^2/Q_{sat}^2(x)$.

It is timely to discuss here the limitation of the space-like kinematics considered in current calculation. In such a case, the wavefunction for an incoming photon with spacelike virtuality $q^2=-Q^2<0$ is very well known and can easily used to estimate the exclusive diffractive cross section, taking $Q^2=M_{\ell^+\ell^-}^2$ for the virtuality of final state photon. Within the color dipole approach, the correct timelike kinematics has been addressed in Ref. \cite{MW} in the scope of exclusive diffractive photoproduction of heavy gauge bosons, like $Z^0$ production. However, a phenomenological study for timelike Compton scattering has not been addressed yet. The main difficulty is the numerical computation of scattering amplitude as the wavefunction for a outgoing photon with timelike $q^2=M_{\ell^+\ell^-}^2>0$ leads to an overlap function which gives an integrand strongly oscilatory as a function of transverse dipole size $r$ as noticed in \cite{MW}. On the other hand, the calculation of timelike Compton scattering  using the $k_{\perp}$-factorization formalism has been done in Ref \cite{antonitcs} as discussed in previous section. It was found the the ratio of the cross section $R$(timelike/spacelike) is $q^2$-dependent and is of order 3 or 4 at high energies.

We notice that the non-forward amplitude is assumed to be partially factorized concerning the $x$ and $t$ dependences, Eqs. (8) and (9), as the saturation scale also depends on the momentum transfer $t$. The model we are using describes correctly the $t$-dependence of vector meson electroproduction and  deeply virtual
Compton scattering  (DVCS) as well.  In general, within the dipole approach the extension of amplitude to finite momentum transfers is set to be completely factorized. This assumption, due to its simplicity, is widely used in the literature, although it contradicts
the Regge-pole theory, as was shown e.g. in studies of DVCS or general parton distributions (GPD), as shown e.g. in Refs. \cite{R1,R2}. We believe that the distinct assumptions for the $t$-dependence will introduce an uncertainty on the overall normalization as we are considering only $t$-integrated cross sections in current study. On the other hand, this issue would be important when distributions on momentum transfer is concerned.

Finally, let us compare the present calculation to previous studies. In Ref. \cite{BCJS} the inclusive dilepton photoproduction has first addressed focusing on the DESY-HERA energy regime. The current calculation does an upgrade for those results using the state of art concerning new parton  distribution functions for the proton and photon. Our predictions for the LHeC are consistent with an energy extrapolation from the HERA machine. However, the LHeC machine is probing the PDFs at smaller values of parton momentum fraction, $x_p$, than for HERA. For instance, for invariant mass of 4 GeV/c, one has the minimum momentum fraction probed $x_p=8\times  10^{-6}$  for LHeC against $x_p=4\times 10^{-4}$ for HERA. Similar trend occurs for the photon parton function as $x_p\rightarrow 1$ and $x_{\gamma}$ goes to $M^2/s$. Therefore, LHeC will be an outstanding place to study the small-$x$ physics in addition to the PDFs for photons.  Concerning the exclusive dilepton photoproduction, our estimates can be compared to recent studies in literature. For instance, in Ref. \cite{antonitcs} the exclusive dilepton photoproduction is computed within the $k_{\perp}$-factorization formalism. We have checked that the invariant mass distribution is consistent with ours with deviations only on the overall normalization. In particular, the extrapolation for the LHeC energy gives an integrated cross section around 1 nb (using the cut $M_{\ell^+\ell^-}^2\geq 1.5$ GeV$^2$/c$^2$). This is very large compared to our estimates and the origin of deviation is the fact that the correct time-like configuration produces a larger cross section compared to the space-like configuration we have considered. This issue is discussed in detail in Ref. \cite{antonitcs} and the main point is that the time-like calculation is a factor 3 or 4 compared to the space-like photon configuration. In addition, the timelike Compton scattering was also investigated in Ref. \cite{PSW} within the generalized parton distributions (GPD) factorization approach. The latter is a robust study as the interference at the amplitude level with the pure QED subprocess is also computed. For the energies we are considering here the QED Bethe-Heitler contribution saturates around 28.5 pb for the integrated cross section whereas the timelike process contributes with 15-25 pb. Disregarding interference contribution (in high energies they are comparable to timelike and QED contributions), the total cross section is around 45-50 pb, which is smaller than the present calculation. The reason for a lower timelike cross section in Ref. \cite{PSW} is the higher order contributions that involve gluon GPDs were not considered.

\section{Summary}
\label{conc}

We have investigated the possibility for the inclusive and exclusive dilepton detection in the photoproduction process at the proposed Deep Inelastic Electron-Nucleon Scattering at the LHC (LHeC) machine. The photon-proton cross sections have been computed for inclusive photoproduction at NLO accuracy and are of order the unit of picobarns. The number of events is evaluated to be of order $10^5$ assuming an integrated luminosity of 10 fb$^{-1}$, which means that such a measurement is fairly feasible. It was shown that that process is a good place to investigate constraints to the parton distribution functions for the proton and the photon at very small-$x$ region.  We have investigated also the exclusive dilepton photoproduction using the LHeC design. It was found that the spectrum is dominated by small invariant masses and the forward amplitude scales with $(Q_{\mathrm{sat}}^2/M^2)$ modulo logarithmic enhancements, where $Q_{\mathrm{sat}}$ is the phenomenological parton saturation scale. Thus, the differential cross section reads as $d\sigma/dM^2\propto (M^2)^{-3}[1+\log(M^2)]$.  The integrated cross section gives $\sigma_{\mathrm{exc}}(\gamma +p\rightarrow \ell^+\ell^-+p)\simeq 80$ pb with the cut $M_{\ell^+\ell^-}\geq 1.5$ GeV/c and also one has  $\sigma(e p\rightarrow \ell^+\ell^-+p)=2.1$ pb for the LHeC energy regime and the number of events is around $1.2\times 10^4$. It can be considered a lower bound for predictions as the timelike photon configuration would give cross sections bigger by a factor three or four.  The number of events is one order of magnitude smaller than for the inclusive case whereas the experimental detection is improved due to the presence of one rapidity gap in final state. 

\begin{acknowledgments}
 This research was supported by CNPq and FAPERGS, Brazil. 
\end{acknowledgments}


\begin{thebibliography}{99}

\bibitem{dainton}
  J.~B.~Dainton, M.~Klein, P.~Newman, E.~Perez and F.~Willeke,
  %``Deep inelastic electron nucleon scattering at the LHC,''
  JINST {\bf 1}, P10001 (2006).

  \bibitem{desreport} J.~L.~Abelleira Fernandez {\it et al.}  [LHeC Study Group Collaboration], J. Phys. {\bf G39}, 075001 (2012).

\bibitem{Magno} M.V.T. Machado, Phys. Rev. D {\bf 78}, 034016 (2008).

\bibitem{CDFmu} A. Aaltonen  {\it et al.} [CDF Collaboration], Phys. Rev. Lett. {\bf 102}, 242001 (2009).

\bibitem{MM} C. Brenner Mariotto and M.V.T. Machado, Phys. Rev. {\bf D86}, 033009 (2012).

\bibitem{BCJS} A.C. Bawa, K. Charchula and W.J. Stirling, Phys. Lett. {\bf B313}, 461 (1993).


\bibitem{dipole}  N.~N.~Nikolaev and B.~G.~Zakharov, Z. Phys. {\bf  C49}, 607
(1991); Z. Phys. {\bf C53}, 331 (1992); A.~H.~Mueller, Nucl. Phys.
{\bf B415}, 373 (1994); A.~H.~Mueller and B.~Patel, Nucl. Phys. {\bf B425}, 471
(1994).


%%
\bibitem{KMW}
H. Kowalski, L. Motyka and G. Watt,
{\it Phys. Rev.} {\bf D74}, 074016 (2006).


\bibitem{MPS}
C.~Marquet, R.~Peschanski and G.~Soyez,
  %``Exclusive vector meson production at HERA from QCD with saturation,''
  Phys.\ Rev.\  D {\bf 76}, 034011 (2007).


\bibitem{antonitcs} W. Sch\"{a}fer, G. \'Slipek and A. Szczurek, Phys.\ Lett.\ B {\bf 688}, 185 (2010). 

%%
\bibitem{Iancu:2003ge}
  E.~Iancu, K.~Itakura and S.~Munier,
  %``Saturation and BFKL dynamics in the HERA data at small x,''
  Phys.\ Lett.\ B {\bf 590}, 199 (2004).


\bibitem{Pumplin:2002vw}
  J.~Pumplin, D.~R.~Stump, J.~Huston, H.~L.~Lai, P.~Nadolsky and W.~K.~Tung,
  JHEP {\bf 0207}, 012 (2002).

\bibitem {grvphoton} M. Gl\"{u}ck, E. Reya and A. Vogt, Phys. Rev. {\bf D45}, 3986 (1992).

\bibitem{MW} L. Motyka and G. Watt, Phys. Rev. {\bf D78}, 014023 (2008).

\bibitem{R1} M. Capua {\it et. al}.,  Phys. Lett. {\bf B645}, 161 (2007).

\bibitem{R2} Laszlo L. Jenkovszky,  Phys. Rev. D74, 114026 (2006).



\bibitem{PSW} B. Pire, L. Szymanowski and J. Wagner, Phys. Rev. {\bf D79}, 014010 (2009).


\end{thebibliography}
\end{document}